\documentclass[preprint,showpacs,preprintnumbers,amsmath,amssymb]{revtex4}

\usepackage{graphicx}
\usepackage{dcolumn}
\usepackage{bm}
\usepackage{colordvi}

\begin{document}

\title{Comparison of covariant and orthogonal Lyapunov vectors}

\author{Hong-liu Yang}
 \email{hongliu.yang@physik.tu-chemnitz.de}
\author{G\"unter Radons}%
 \email{radons@physik.tu-chemnitz.de}
\affiliation{%
Institute of Physics,
Chemnitz University of Technology, D-09107 Chemnitz, Germany 
}

\date{\today}
\begin{abstract}
Two sets of vectors, covariant and orthogonal Lyapunov vectors (CLVs/OLVs), are currently used to characterize the linear 
stability of chaotic systems. A comparison is made to show their similarity and difference, especially 
with respect to the influence on hydrodynamic Lyapunov modes (HLMs). 
Our numerical simulations show that in both Hamiltonian and dissipative systems HLMs formerly detected via
OLVs survive if CLVs are used instead. Moreover the previous classification of two universality classes works for CLVs as well, i.e. the dispersion relation is linear for
Hamiltonian systems and quadratic for dissipative systems respectively. 
The significance of HLMs changes in different ways for Hamiltonian and dissipative systems with the replacement
of OLVs by CLVs. For general dissipative systems with nonhyperbolic dynamics the long wave length structure in Lyapunov vectors corresponding to near-zero Lyapunov exponents 
is strongly reduced if CLVs are used instead, whereas for highly hyperbolic dissipative systems the significance of HLMs is nearly identical for CLVs and OLVs.
In contrast the HLM significance of Hamiltonian systems is always comparable for CLVs and OLVs irrespective of hyperbolicity. 
We also find that in Hamiltonian systems different symmetry relations between conjugate pairs are observed for CLVs and OLVs. Especially, CLVs in a conjugate pair 
are statistically indistinguishable in consequence of the micro-reversibility of Hamiltonian systems. Transformation properties of Lyapunov exponents, CLVs and hyperbolicity
under changes of coordinate are discussed in appendices. 
\end{abstract}


\maketitle

\section{Introduction}
Chaos means a sensitive dependence on initial conditions. This intrinsic randomness of the fully deterministic systems makes a statistical
treatment of them feasible, which is essential for the foundations of statistical mechanics \cite{dorfman}.
 Besides, chaos plays an important role in a plenty
of phenomena which is of relevance to our daily life, for instance the weather forecasting \cite{weather}.

To characterize the chaoticity of dynamical systems, Lyapunov exponents and vectors are mostly used. One important recent finding
of Lyapunov analysis is that for systems with continuous symmetries Lyapunov vectors corresponding to near-zero Lyapunov exponents have long wave-length
structures, named hydrodynamic Lyapunov modes (HLMs) \cite{posch-hirschl}. This provides a new possibility to connect the reduced description of a many-body system to the microscopic information of
 its detailed dynamics. Further investigations showed that HLMs exist in a large number of systems \cite{eckmann,france,yang,hlm-cml} and they have some
universal features irrespective of the details of their dynamics \cite{hlm-univ}. One should mention that localization of Lyapunov vectors corresponding to the
largest Lyapunov exponents was also intensively studied \cite{pikovsky2}.

Lyapunov analysis was conventionally undertaken via the so-called Benettin algorithm \cite{benettin}, where 
Lyapunov vectors are calculated as the set of orthogonal vectors right after reorthogonalization of offset vectors. 
Recently, the application of another set of vectors called covariant
Lyapunov vectors (CLVs) was made feasible via an efficient algorithm proposed by Ginelli et al.\cite{clv}.
CLVs have been shown suitable for the characterization of hyperbolicity of high dimensional systems since
they are expected to span the local stable and unstable subspaces of the investigated systems. 
In view of the obvious difference, it becomes necessary to study the relation between CLVs and the conventionally used Lyapunov vectors. 
We denote the latter as orthogonal Lyapunov vectors (OLVs) in order to distinguish them from CLVs.

We first recall in Sec.II the definition and numerical calculation of both sets of vectors. The model system of coupled map
lattices (CMLs) is introduced in Sec. III. Through intensive numerical simulations the following questions are addressed in the remaining
sections: (i) will HLMs survive if CLVs are used instead of OLVs (Sec. IV), (ii) are HLMs from CLVs as significant as those from OLVs,
(iii) what is the implication of the Hamiltonian structure to the relation between conjugate pair of CLVs, and what is the implication for
the relation between coordinate and momentum parts of CLVs?

\section{definition and calculation algorithm for CLVs and OLVs} \label{sec:algorithm}
Recall that in the seminal work \cite{osel} about the multiplicative ergodic theorem Oseledec proved that the limit 
$\Xi =\lim_{t \to +\infty} [ M^T(t,0)\cdot M(t,0)]^{\frac{1}{2t}}$
exists for almost every initial point of a nonlinear dynamical system, where
$M(t,0)$ is the fundamental matrix governing the time evolution of perturbations 
$\delta X(t)$ in tangent space as $\delta X(t) =M(t,0) \cdot \delta X(0)$. 
The set of Lyapunov exponents are defined as $\lambda^{(\alpha)} =\ln \mu^{(\alpha)}$, where $\mu^{(\alpha)}$ are 
the eigenvalues of the matrix $\Xi$,
 i.e. $\Xi \cdot \bf{g}^{(\alpha)} =\mu^{(\alpha)} \bf{g}^{(\alpha)}$.

In practice the Lyapunov exponents and OLVs are calculated via the so-called standard method 
invented by Benettin and Shimada et al. \cite{benettin}, which was used in most studies of HLMs \cite{posch-hirschl,eckmann,france,yang,hlm-cml,hlm-univ}. 
Here the time evolution of a set of offset vectors in tangent space is monitored by integrating the linearized equation. 
And the offset vectors are reorthonormalized periodically. 
The time averaged values of the logarithms of the renormalization factors are
the Lyapunov exponents and the set of offset vectors $\bf{f}^{(\alpha)}$ right 
after the reorthonormalization are the Lyapunov vectors. 
The relation between the Oseledec 
eigenvectors $\bf{g}^{(\alpha)}$ and the Lyapunov vectors obtained via the standard method is subtle. 
It is proved that the Lyapunov vectors $\bf{f}^{(\alpha)}$ obtained via the standard 
method converge exponentially to the Oseledec 
eigenvectors for the inverse-time dynamics of the original system \cite{goldhirsch,ershov}. In other words, as $t \to +\infty$,
there is $\bf{f}^{(\alpha)} \sim \overline{\bf{g}}^{(\alpha)}$, where $\overline{\bf{g}}^{(\alpha)}$ are eigenvectors of the matrix
$\overline{\Xi}=\lim_{t \to +\infty} [ \overline{M(t,0)}^T\cdot  \overline{M(t,0)}]^{\frac{1}{2t}}$
as well as its inverse
$\overline{\Xi}^{-1}=\lim_{t \to +\infty} [ M(t,0)\cdot  M(t,0)^T]^{\frac{1}{2t}}$  and $\overline{M(t,0)} \equiv [M(t,0)]^{-1}$
is the fundamental matrix of the inverse-time dynamics.
See Ref.\cite{goldhirsch,ershov} for the details.

Exactly in the same work \cite{osel} Oseledec proved also that, for almost all initial conditions $x$, 
there is a splitting of the tangent space $TM(x)$
\begin{equation}
\label{sp}
TM(x)=E^{1}(x) \oplus E^2(x)\oplus \cdots \oplus E^s(x),
\end{equation}
and there exist real numbers $\lambda_1(x)>\lambda_2(x)>\cdots >\lambda_s(x)$ such that 
\begin{equation}
\lim_{n \to \pm \infty} \frac{1}{n} \ln \partial Df^n|_{E^i} \partial=\lambda_i(x),
\end{equation}
where $Df$ is the derivative governing the tangent space dynamics. 
The set of numbers $\lambda_i(x)$ with degeneracy $m_i={\text{dim}} E^i(x)$ composes the Lyapunov spectrum 
and the decomposition stated in
Eq.(\ref{sp}) is called the Oseledec splitting. The spanning vectors $ \bf{e}^{(\alpha)}$ of the Oseledec subspace $E^i(x)$ are the CLVs.

In contrast to the popularity of OLVs, the use of CLVs was made feasible only recently owing to an efficient algorithm
 proposed by Ginelli et al. \cite{clv}. 
The new algorithm relies on the information obtained via the standard method of Benettin. 
One additional integration of the inverse-time
dynamics is performed in order to get CLVs and the corresponding fluctuating finite-time Lyapunov exponents.
The basic idea is that an arbitrary offset vector will approach asymptotically the most unstable direction corresponding to the largest
Lyapunov exponent. 
It is known that the covariant $k$-dimensional subspace spanned by the first $k$ CLVs is spanned by the first
$k$ OLVs as well. An arbitrary offset vector confined to this subspace will approach asymptotically the $k$-th CLV if 
the inverse-time tangent space dynamics is
applied. To this aim one needs the effective tangent space dynamics confined in the $k$-dimensional covariant subspace. A representation
of this effective dynamics in the coordinate space of OLVs is given by the R-matrix produced by the reorthonormalization steps of the
standard method. Detailed formulas can be found in Ref. \cite{clv}. 

We mention that a different algorithm was used in Ref. \cite{clv-local}. It is demanded to compare the efficiency of the two algorithms.

To characterize Lyapunov vectors of extended systems quantitatively, we
introduced in \cite{yang} a dynamical variable called {\it LV fluctuation density} in the spirit of generalized hydrodynamics, 
\begin{equation}
{\cal U}^{(\alpha)}(r,t)=\sum_{l=1}^L \delta u^{(\alpha)l}_t \cdot \delta(r-r_l(t)),
\end{equation}
where $\delta (x)$ is Dirac's delta function, $r_l(t)\equiv l\cdot a$ is the position coordinate of the $l$-th element taken here as 
$r_l(t)\equiv l\cdot a$, 
and $\{\delta u^{(\alpha)l}_t\}$ is the coordinate or momentum part of the $\alpha$-th Lyapunov vector at the discrete time t.
 For simplicity, we set $a=1$ in the following discussion.
The spatial structure of LVs is characterized by the {\it static LV structure factor} defined as
 \begin{equation}
 \label{sk}
S_u^{(\alpha\alpha)}(k)=\int \langle {\cal U}^{(\alpha)}(r,0)  {\cal U}^{(\alpha)}(0,0) \rangle   e^{-jk\cdot r}dr,
\end{equation}
which is just the spatial power spectrum of the LV fluctuation density. 

As shown in past studies \cite{yang,hlm-cml,hlm-univ}, the quantity $k_{max}$ representing the wave-number of the 
dominant peak of the structure factor $S(k)$ and $S(k_{max})$ can be used to characterize the significance of long wave length
structure in Lyapunov vectors.

\section{models}
CMLs \cite{cml} were selected as the main focus of this study because they have, 
which is essential to HLMs, similar symmetries as many-particle systems but are relatively much simpler.                    
  
 The two classes of CMLs under investigation have the form
\begin{subequations}
\label{map-standard}
\begin{equation}
\label{map-standard-a}
v_{t+1}^l= v_t^l+\epsilon [ f(u_t^{l+1}-u_t^l)-f(u_t^l-u_t^{l-1})]
\end{equation}
\begin{equation}
u_{t+1}^l=u_t^l+v^l_{t+1}
\end{equation}
\end{subequations}
and 
\begin{equation}
\label{map-circle}
u_{t+1}^l= u_t^l+\epsilon [ f(u_t^{l+1}-u_t^l)-f(u_t^l-u_t^{l-1})]
\end{equation}
where $f(z)$ is a nonlinear map, $t$ is the discrete time index, $l=\{1,2,\cdots,N\}$ is
the index of the lattice sites and $N$ is the system size. Unless explicitly stated,
we use periodic boundary conditions
$\{u_t^0=u^N_t,u_t^{N+1}=u_t^1\}$ in the numerical simulations below. 

Two options of the local map are used,
the sinusoidal map $f_{C}(z)=\frac{1}{2\pi}\sin(2\pi z)$ and the skewed tent map
\begin{equation}
\label{map-sktent}
f_{T}(z)=
\begin{cases}
 z'/r  &   \text{for $0<z'\leq r$},  \\
 (1-z')/(1-r)  & \text{for $r<z'<1$.   }
\end{cases}
\end{equation}
with $z'=z \pmod 1$. 
With the parameter being close to zero Eqs. (\ref{map-standard}) and (\ref{map-circle}) with the skewed tent map is highly hyperbolic whereas Eqs. (\ref{map-standard}) and
(\ref{map-circle}) with the sinusoidal map is nonhyperbolic. Especially the Hamiltonian system Eq. (\ref{map-standard}) with the two options of the local map are similar to the
well-studied cases of hard-core systems and soft-potential systems, respectively. Alternatively, tuning the parameter $r$ of Eq. (\ref{map-sktent}) 
from $0$ to $0.5$ leads to a smooth variation of the dynamics of Eq. (\ref{map-standard}) from hard-core-like
to soft-potential like \cite{hlm-when}.  

Obviously, both systems Eq. (\ref{map-standard}) and (\ref{map-circle}) are invariant under an arbitrary translation in $u$-direction. 
Such a symmetry is known to be responsible for the appearance of HLMs.

\section{existence of HLMs in CLVs} 
\begin{figure}
\includegraphics*[scale=0.35]{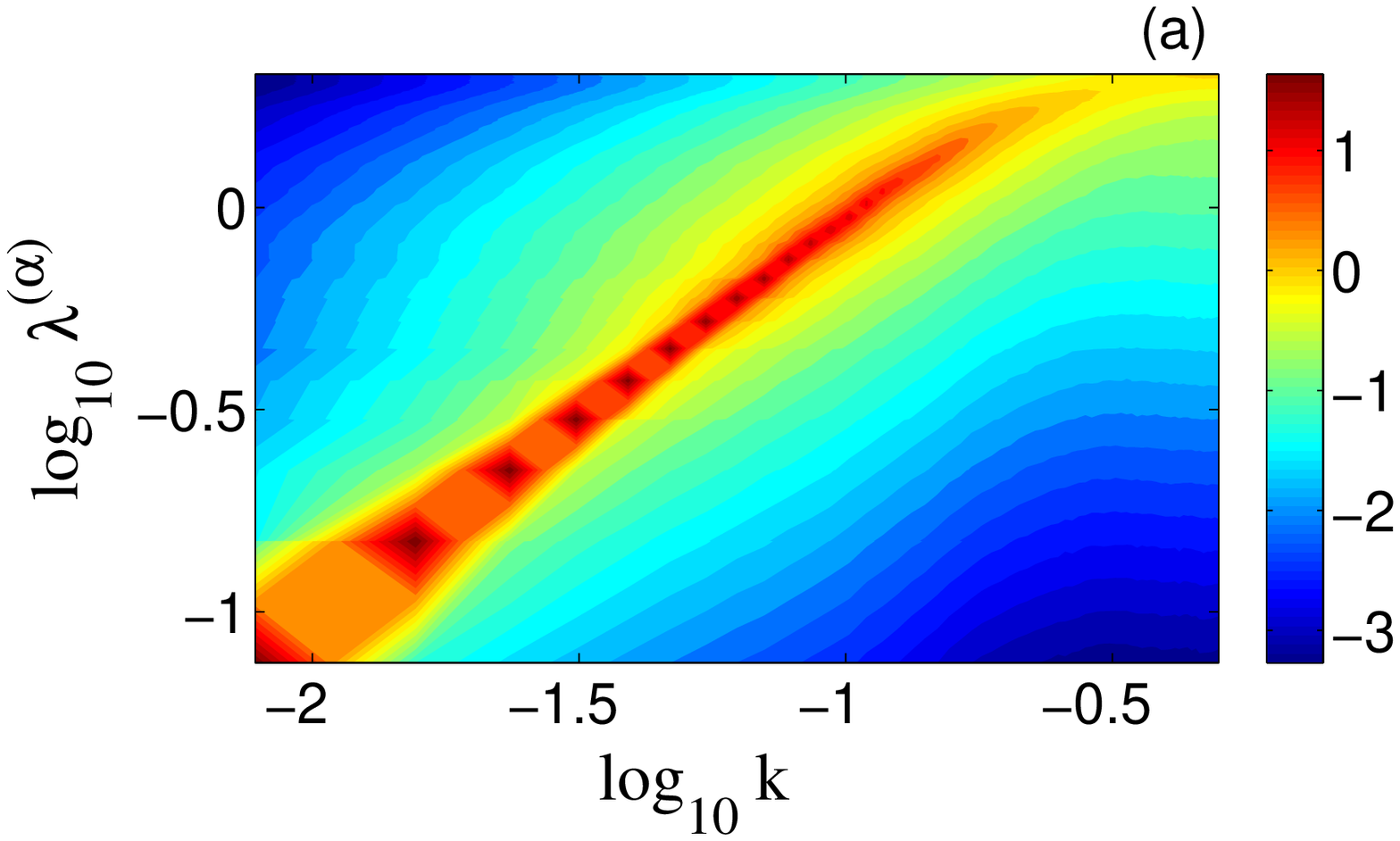}\\
\includegraphics*[scale=0.35]{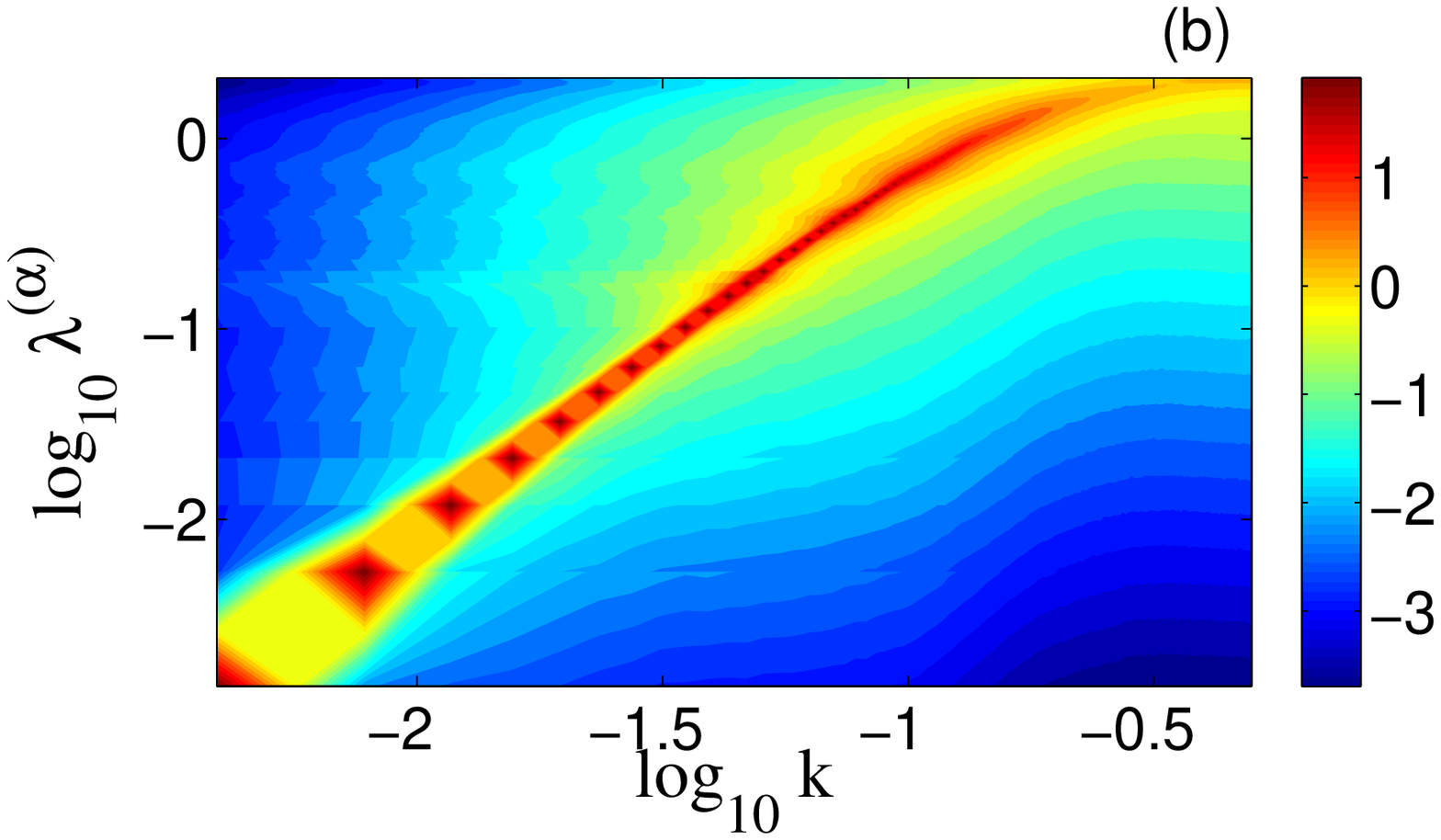}\\
\includegraphics*[scale=0.35]{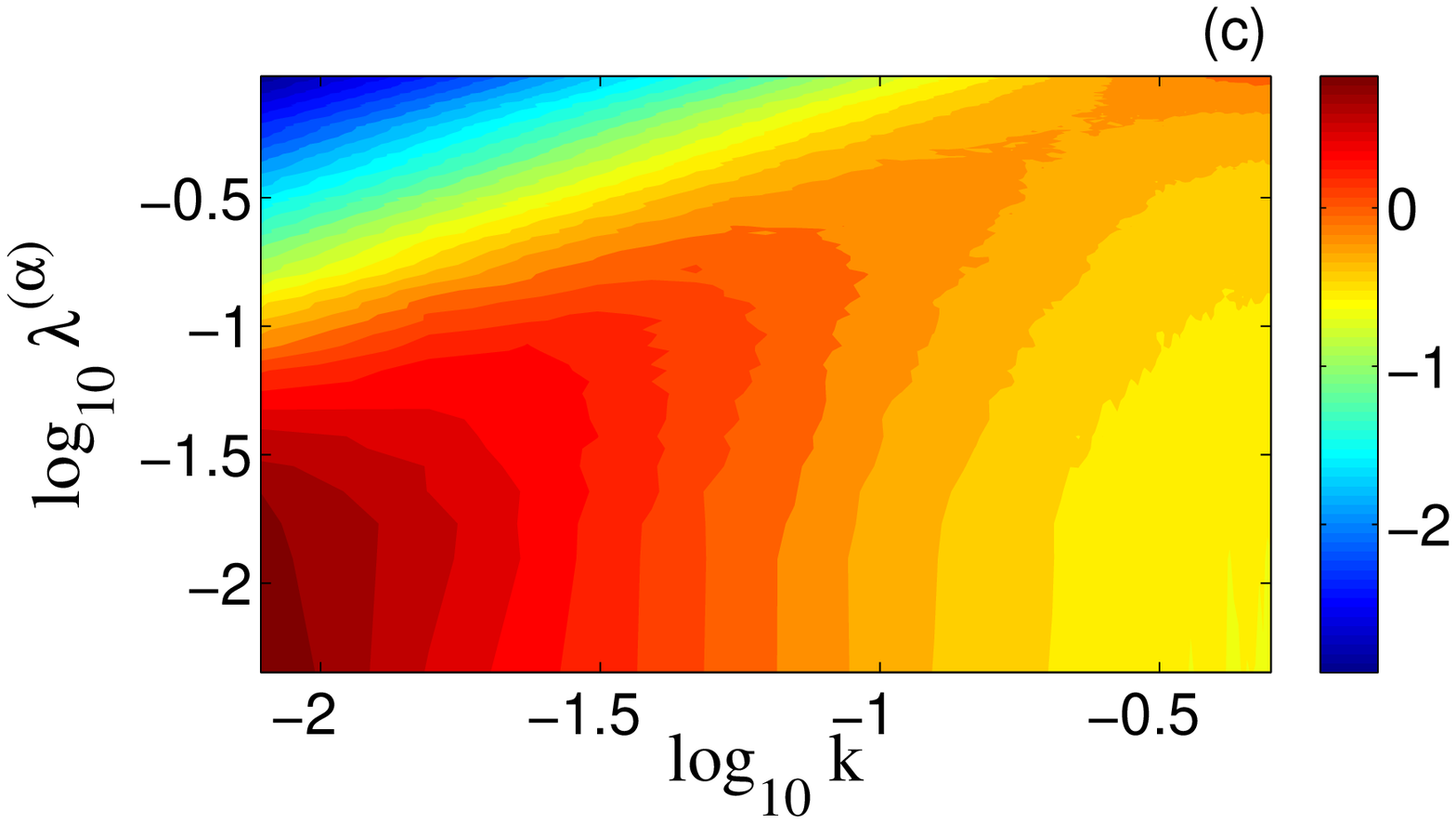}\\
\includegraphics*[scale=0.35]{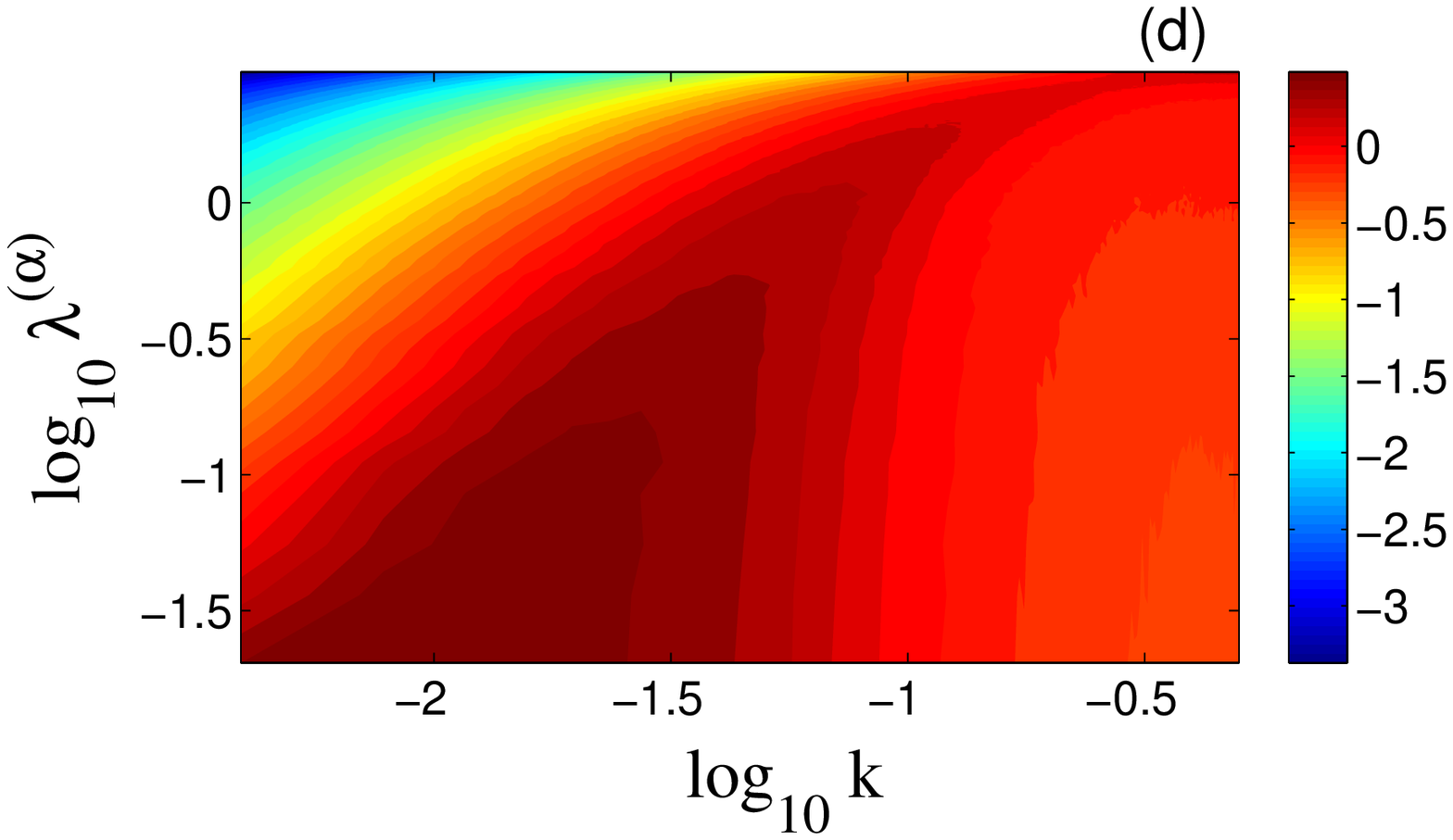}
\caption{\label{fig:f1}
Contour plot of the static CLV structure factors for (a,c) Hamiltonian and (b,d) dissipative coupled map lattices. The local dynamics used
is the skewed tent map (a,b) and the sinusoidal map (c,d), respectively. Other parameters are $\epsilon=1.3$ and $r=0.15$. A ridge structure can be clearly
seen in the small $(\lambda,k)$ regime which indicates the existence of HLMs.
}
\end{figure}

We show in Fig.\ref{fig:f1} the contour plot of the static CLV structure factors $S(k)$. For both dissipative and Hamiltonian systems, either with the skewed tent map or the
sinusoidal map, a clear ridge structure can be seen in the regime $(k,\lambda) \sim (0,0)$, which indicates the existence of long wave-length structures in CLVs
 associated with near-zero Lyapunov exponents. 
These numerical results demonstrate that HLMs formerly detected via OLVs survive if CLVs are used instead.

\section{universality of dispersion relations} 
In Ref.\cite{hlm-cml,hlm-univ} we found that the $\lambda$-$k$ dispersion relation of HLMs can be classified into two universality classes with respect to the system dynamics. 
Dissipative systems have a quadratic $\lambda$-$k$ dispersion while Hamiltonian systems have a linear one. Now we see whether such a classification is still valid if CLVs are used
instead.  
 
\begin{figure}
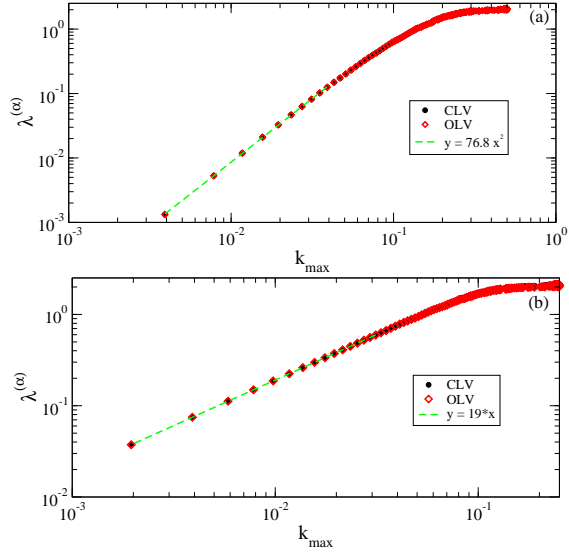

\includegraphics*[scale=0.3]{disp-dissip.eps}\\
\includegraphics*[scale=0.3]{disp-ham.eps}
\caption{\label{fig:f7}
Dispersion relations $\lambda$-$k_{max}$ obtained from CLVs and OLVs for (a) dissipative and (b) Hamiltonian systems, respectively. 
The local map is the skewed tent map with
$\epsilon=1.3$ and $r=0.15$. Note the perfect agreement between data from CLVs and OLVs for the highly hyperbolic cases shown here.
}
\end{figure}

The cases with the skewed tent map as local dynamics are shown in Fig.\ref{fig:f7}. As can be seen from the plot, the CLV dispersion relations for dissipative and
Hamiltonian systems have the different asymptotic behavior. The former is of the asymptotic form $\lambda \sim k_{max}^2$ while the latter is $\lambda \sim k_{max}$, as reported for OLVs \cite{hlm-cml}.
Moreover, for the used parameter setting the dispersion curves for CLVs agree very well with those of OLVs .
\begin{figure}
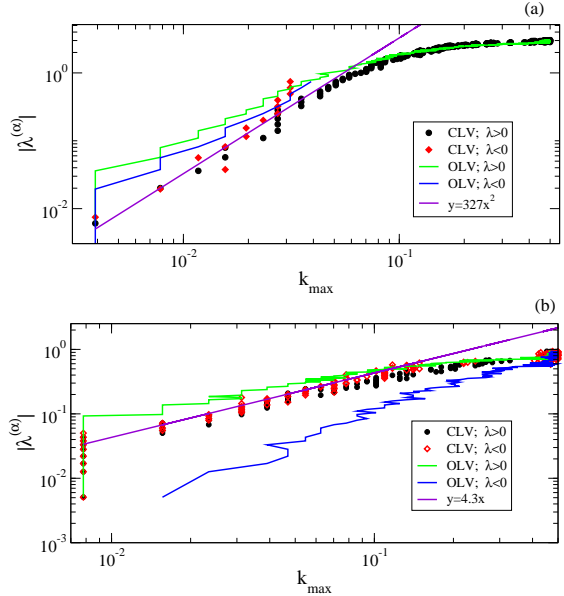

\includegraphics*[scale=0.3]{disp-1d-stan-4.eps}\\
\includegraphics*[scale=0.3]{disp-ham-stan.eps}
\caption{\label{fig:f8}
Similar to Fig. \ref{fig:f7} but the local map is the sinusoidal map and $\epsilon=1.3$. Note that for the nonhyperbolic cases shown in this figure
CLVs and OLVs still have the same asymptotic behavior.
}
\end{figure}
Cases with the sinusoidal map as local dynamics are shown in Fig.\ref{fig:f8}. For both dissipative and Hamiltonian systems CLV dispersions are converging 
to the expected asymptotic forms, even better than OLVs. 
Note also that, as shown in Fig.\ref{fig:f8}b, for Hamiltonian systems 
CLV dispersions for positive and negative Lyapunov exponents follow the same curve while OLV dispersions behave differently in the positive and negative Lyapunov exponent
regimes. For the used system size only the positive Lyapunov exponent branch of OLV dispersion is close to the asymptotic form. Further discussion regarding these differences
 will be given in
the following sections. Nevertheless, for the two representative cases the investigated CLV dispersions follow well the reported classification of the universality classes of 
HLMs \cite{hlm-cml,hlm-univ}.

\section{significance of HLMs} 
To characterize the significance of long wave length structure in Lyapunov vectors, we use the measure $S(k_{max})$, which is the height of the dominant peak in the static
LV structure factor $S(k)$ (Eq. (\ref{sk}).

\begin{figure}
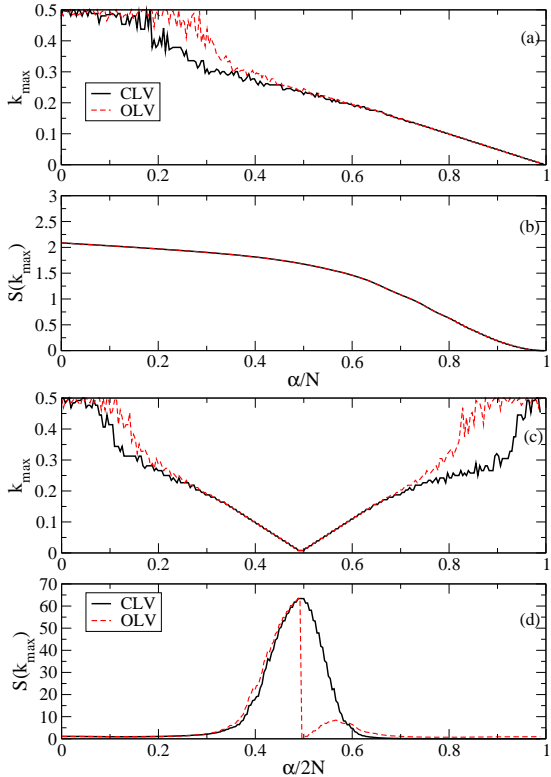

\includegraphics*[scale=0.3]{peak-clv-gslv-dissip-tent.eps}\\
\includegraphics*[scale=0.3]{peak-clv-gslv-ham-sktent.eps}
\caption{\label{fig:f2}
Dominant wave number $k_{max}$  and the significance measure $S(k_{max})$ of
 CLVs and OLVs for (a,b) dissipative and (c,d) Hamiltonian systems respectively. The local map is the skewed tent map with
$\epsilon=1.3$ and $r=0.15$. The long wave length structure is as significant in CLVs as in OLVs for the highly hyperbolic cases shown. 
}
\end{figure}

\begin{figure}
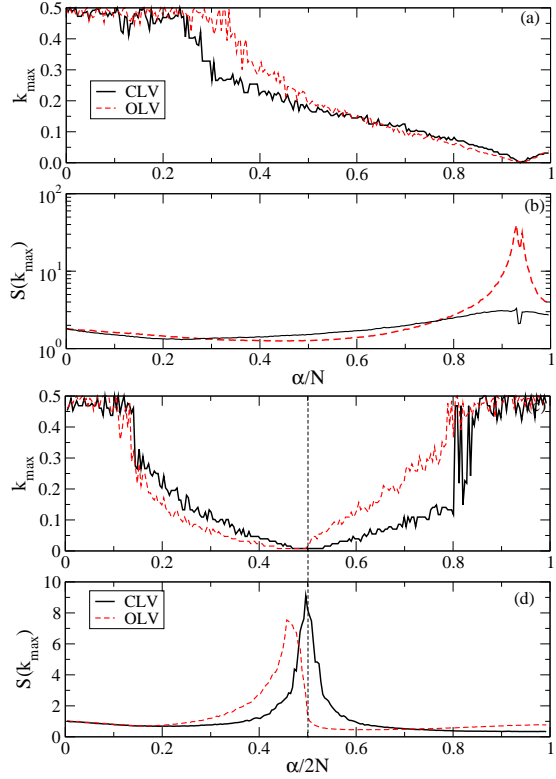

\includegraphics*[scale=0.3]{peak-clv-gslv-dissip-stan-4.eps}\\
\includegraphics*[scale=0.3]{peak-clv-gslv-ham-stan.eps}
\caption{\label{fig:f3}
Similar to Fig. \ref{fig:f2} but the local map is the sinusoidal map and $\epsilon=1.3$. The significance of HLM is strongly reduced in CLVs of nonhyperbolic dissipative systems
whereas the HLM significance is comparable for CLVs and OLVs in nonhyperbolic Hamiltonian systems.
}
\end{figure}

The dominant wave number $k_{max}$ and the significance measure $S(k_{max})$ are compared for CLVs and OLVs in Fig.\ref{fig:f2} for the cases with the skewed tent map as the
local dynamics. For such highly hyperbolic systems both the position and the height of the dominant peak are nearly identical for CLVs and OLVs
for either the dissipative system or the Hamiltonian system in the positive Lyapunov exponent regime. We postpone the discussion of the negative Lyapunov exponent part of 
the Hamiltonian system to the next section. This observation indicates that for the highly hyperbolic systems the significance of HLMs is not influenced if CLVs are used instead
of OLVs.
A similar comparison was made also for cases with the sinusoidal map as the local dynamics as shown in Fig. \ref{fig:f3}. 
For both, dissipative system and Hamiltonian system, clear
discrepancies between CLVs and OLVS can be seen in $k_{max}$ and $S(k_{max})$. For the dissipative case, 
the height of the dominant peak $S(k_{max})$ for CLVs is much lower than for OLVs (see Fig.\ref{fig:f3}), especially in
the regime $\lambda \sim 0$ ($\alpha/N \simeq 0.93$), which means 
that for the strongly nonhyperbolic systems as shown here, the use of CLVs reduces the visibility of long
 wave length structure as compared to OLVs. In contrast, for the Hamiltonian case,
the height of the dominant peak $S(k_{max})$ is comparable for CLVs and OLVs. Note, however, that the variation of $S(k_{max})$ for CLVs is symmetric with respect to the
spectral center
$\alpha/2N=0.5$ while it is asymmetric for OLVs (Fig.\ref{fig:f3}d).

\begin{figure}
\includegraphics*[scale=0.5]{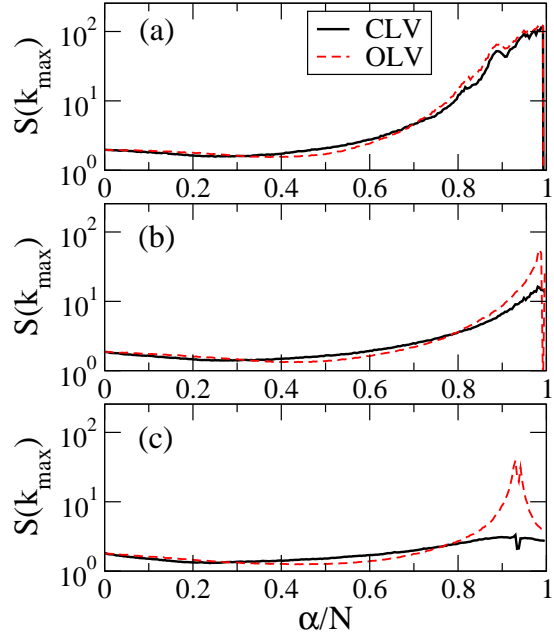}
\caption{\label{fig:f4a}
Influence of the hyperbolicity variation on the significance measure $S(k_{max})$ of HLMs in the dissipative system. The parameter $r$ is
(a) 0.2, (b) 0.3 and (c) 0.4 respectively, which corresponds to a decreasing hyperbolicity.
The local map is the skewed tent map and $\epsilon=1.3$.
}
\end{figure}

\begin{figure}
\includegraphics*[scale=0.5]{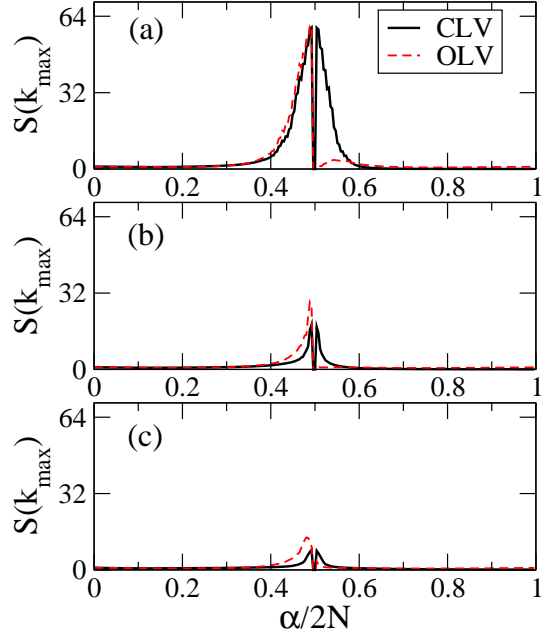}
\caption{\label{fig:f4b}
Influence of the hyperbolicity variation on the significance measure $S(k_{max})$ of HLMs in the Hamiltonian system. The parameter $r$ is
(a) 0.2, (b) 0.3 and (c) 0.4 respectively, which corresponds to a decreasing hyperbolicity.
The local map is the skewed tent map and $\epsilon=1.3$.
}
\end{figure}

To demonstrate further the influence of hyperbolicity on the significance difference between CLVs and OLVs we tune the parameter $r$ of the skewed tent map Eq.(\ref{map-sktent}).
Results for dissipative cases and Hamiltonian cases are shown in Fig. \ref{fig:f4a} and \ref{fig:f4b} respectively. In consistence with our previous results in Ref.
\cite{hlm-when} the weakening of hyperbolicity as increasing $r$ from $0.2$ to $0.4$ leads to a dramatic reduction of the significance of HLMs, in both CLVs and OLVs, for
either dissipative system or Hamiltonian system. In the dissipative
system the reduction of CLV
significance as increasing $r$ is much faster than the reduction of OLV significance, which leads to an increasing discrepancy between them.
 In contrast for the Hamiltonian system
the significance of CLVs and OLVs is always comparable. Note, however, that the tuning of $r$ has no influence on the symmetry features of Lyapunov vectors.

\begin{figure}
\includegraphics*[scale=0.3]{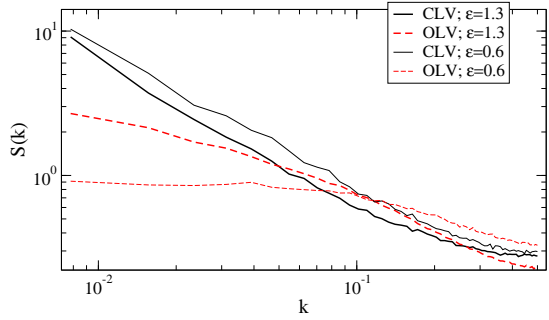}
\caption{\label{fig:f5}
Static LV structure factor corresponding to the smallest positive Lyapunov exponent of the Hamiltonian case. 
The local map is the sinusoidal map. The large difference between these specific CLVs and OLVs is due to the asymmetric locating of the largest value of $S(k_{max})$ of OLVs as shown
in Fig. \ref{fig:f3}b. 
}
\end{figure}

Owing to the different symmetry properties of CLVs and OLVs of Hamiltonian system, the largest $S(k_{max})$ is observed at different $\alpha$
values. This leads to a rather
 large
difference in the static LV structure factors corresponding to the smallest positive Lyapunov exponents. We show in Fig. \ref{fig:f5} two cases with different coupling
strength $\epsilon$. As can be seen from the figure the CLV structure factor $S(k)$ diverges quickly as $k$ goes to zero while the OLV structure factor increases relatively 
slowly and even
seems to saturate to a constant. Note that the same parameter $\epsilon=0.6$ was used in Ref. \cite{clv} (see Fig.3 therein). 

\begin{figure}
\includegraphics*[scale=0.3]{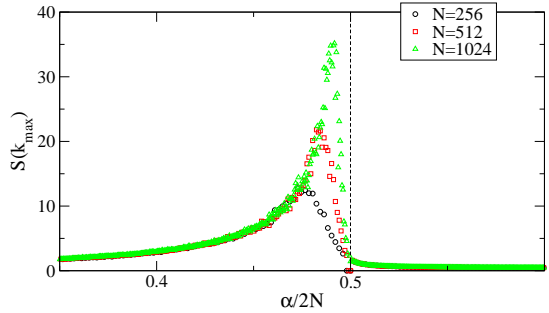}
\caption{\label{fig:f6}
Variation of the significance measure $S(k_{max})$ of HLMs with the system size $N$ for the Hamiltonian case with the sinusoidal map. The peak moves gradually to the spectral center point
$\alpha/2N=0.5$ as increasing $N$.
}
\end{figure}

Moreover, as increasing the system size $N$ the asymmetrically located peak of $S(k_{max})$ for OLVs shifts towards the spectral center point $\alpha/2N=0.5$ as shown in Fig. \ref{fig:f6}, which indicates a
gradual reduction of the discrepancy between CLVs and OLVs as approaching the thermodynamic limit.

\section{conjugate pair relation in Hamiltonian system}
\label{sec:pair}
\begin{figure}
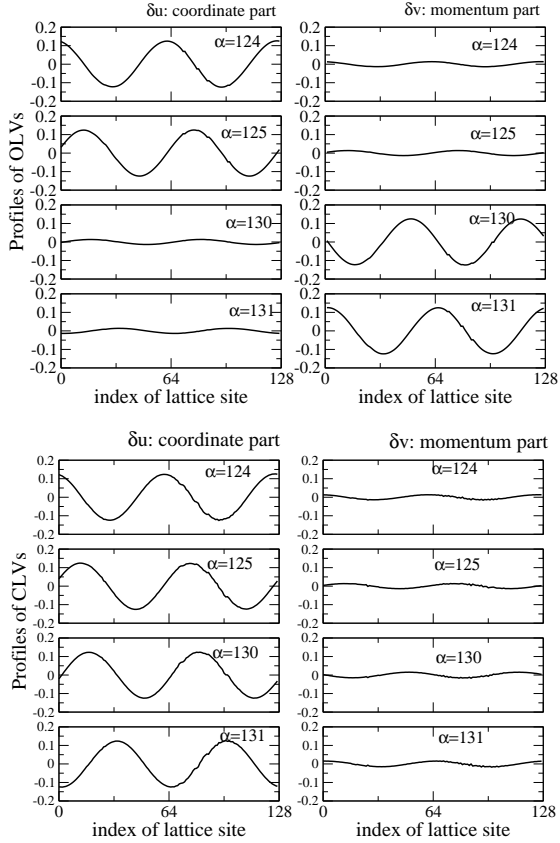

\includegraphics*[scale=0.3]{gslv-profiles-time3.eps}\\
\vskip 0.3cm
\includegraphics*[scale=0.3]{clv-profiles-time3.eps}
\caption{\label{fig:f9}
Instantaneous profiles of two conjugate pairs of CLVs and OLVs for the Hamiltonian case. The local map is the skewed tent map with
$\epsilon=1.3$ and $r=0.15$. The system size used is $N=128$. Note that OLVs and CLVs have different symmetry relations for the conjugate pair. 
}
\end{figure}

By definition, CLVs are in general not mutually orthogonal as OLVs are. This difference has some interesting consequences in Hamiltonian systems.
As reported in Ref.\cite{france,hlm-cml}, a conjugate pair of OLVs with $\lambda^{(\alpha)}=-\lambda^{(2L-1-\alpha)}$ has the symmetry that
 $\delta u^{(\alpha)}=\pm\delta v ^{(2L-1-\alpha)}$ and $\delta v^{(\alpha)}=\mp \delta u ^{(2L-1-\alpha)}$. 
 Here $\delta u$ and $\delta v$ denote the coordinate and momentum parts of LVs respectively.
 The physical origin of this symmetry lies in the symplectic structure of Hamiltonian system. OLVs as the eigenvectors of the matrix
  $\overline{\Xi}=\lim_{t \to +\infty} [ \overline{M(t,0)}^T\cdot  \overline{M(t,0)}]^{\frac{1}{2t}}$ are thus forced to have the observed symmetry.
  Examples of conjugate pairs of OLVs are shown in Fig. \ref{fig:f9}.
 
\begin{figure}
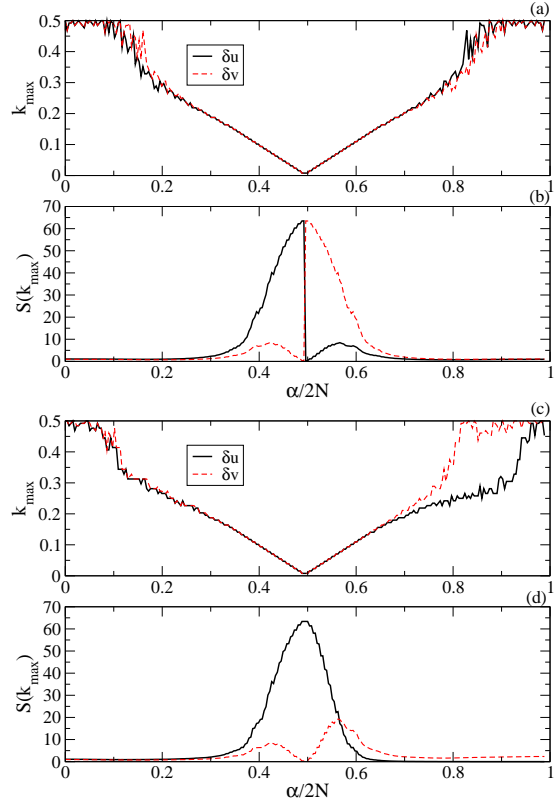

\includegraphics*[scale=0.3]{peak-gslv-dxdp-ham-sktent.eps}\\
\includegraphics*[scale=0.3]{peak-clv-dxdp-ham-sktent.eps}
\caption{\label{fig:f10}
Comparison of the dominant wave number $k_{max}$ and $S^{(\alpha )}(k_{max})$ obtained from the coordinate and momentum parts of OLVs (a,b) 
and CLVs (c,d)
respectively. The local map of the studied Hamiltonian system is the skewed tent map with
$\epsilon=1.3$ and $r=0.15$.
}
\end{figure}

 As can be seen from the same plot, CLVs behave differently. The relation $\delta u^{(\alpha)}=\pm \delta u ^{(2L-1-\alpha)}$ and 
 $\delta v^{(\alpha)}=\mp \delta v ^{(2L-1-\alpha)}$ seems to work well instead. Note also that for CLVs the amplitude of the wave structure 
 in the momentum part is much smaller than
 in the corresponding coordinate part. Such difference is also reflected in the profiles of $S(k_{max})$ in Fig. \ref{fig:f10}. For OLVs $S(k_{max})$ of the
 coordinate part and momentum part are mutual mirror images with respect to the spectral center $\alpha/2N=0.5$. For CLVs $S(k_{max})$ from either the
 coordinate part or the momentum part is roughly symmetric with respect to the center by itself. 
 
\begin{figure}
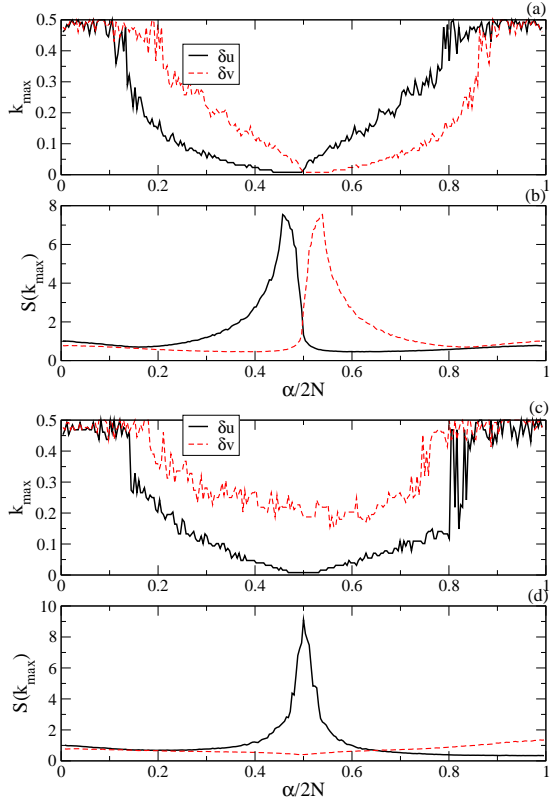

\includegraphics*[scale=0.3]{peak-gslv-dxdp-ham-stan.eps}\\
\includegraphics*[scale=0.3]{peak-clv-dxdp-ham-stan.eps}
\caption{\label{fig:f13}
Similar to Fig. \ref{fig:f10} but the local map is the sinusoidal map and
$\epsilon=1.3$.
}
\end{figure}

 As going to the nonhyperbolic cases with the sinusoidal map as the local dynamics, the mentioned simple relations between $\delta u$ and $\delta v$ of instantaneous Lyapunov
 vectors valid no longer. However, as can be seen from Fig. \ref{fig:f13}, the
 conjugate pairs of CLVs have nearly identical $k_{max}$ and $S^{(\alpha )}(k_{max})$, i.e. they are statistically indistinguishable. 
 This interesting feature of CLVs is believed
 coming from the micro-reversibility of Hamiltonian system. Micro-reversibility means that for each trajectory from $(u(0),v(0))$ to $(u(T),v(T))$ there exists a reverse-time 
 trajectory from $(u(T),-v(T))$ to $(u(0),-v(0))$. Under time reversal Lyapunov exponents change their sign and the conjugate pair of CLVs exchange their role for characterizing the stable
 and unstable directions. Owing to the ergodicity of Hamiltonian system \cite{note-ergocicity} 
 $S^{(\alpha )}(k_{max})$ for the pair of initial
 conditions $(u(0),v(0))$ and $(u(T),-v(T))$ are indistinguishable and this leads to the observed symmetry feature of CLVs.

\begin{figure}
\includegraphics*[scale=0.5]{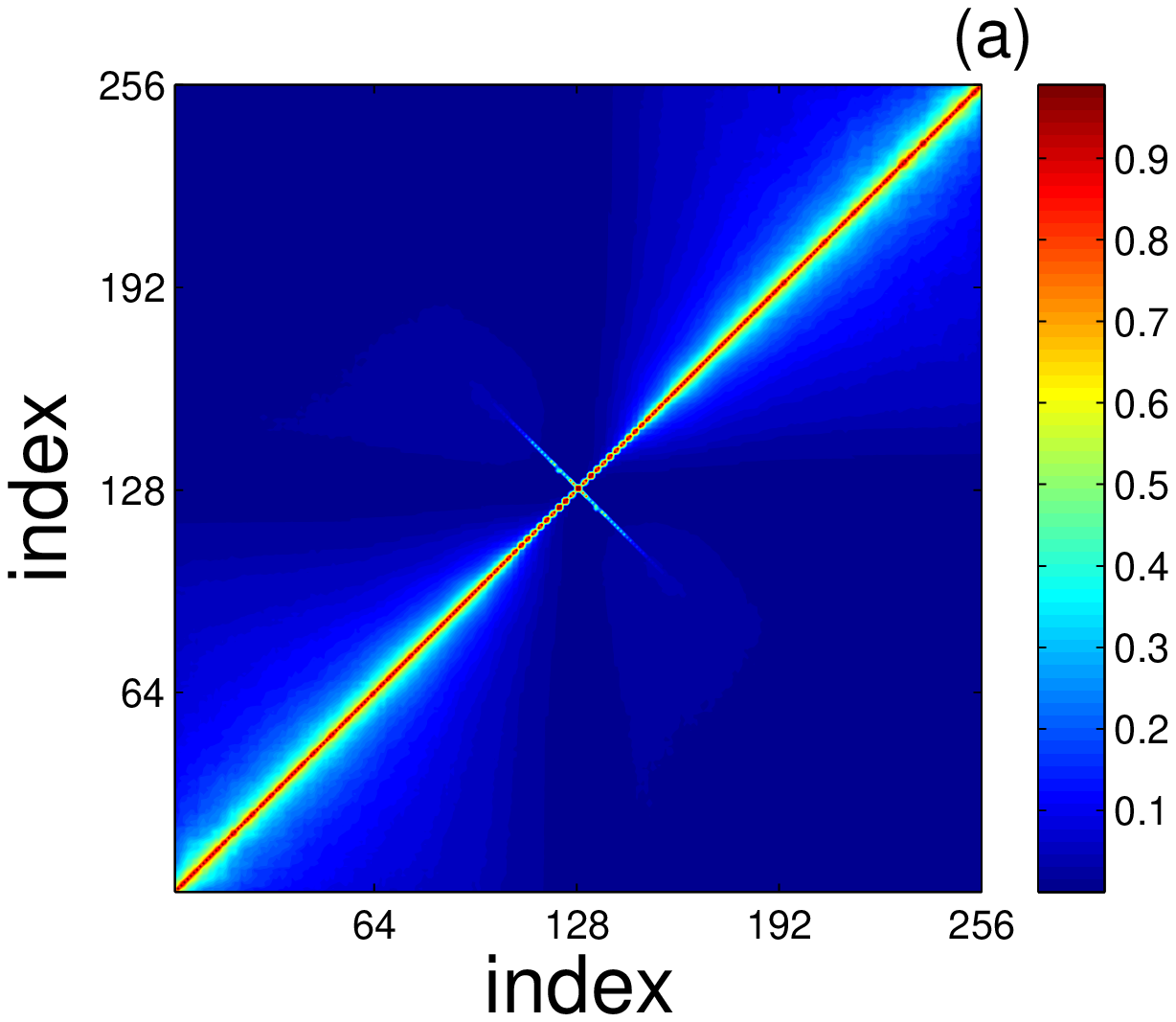}\\
\includegraphics*[scale=0.5]{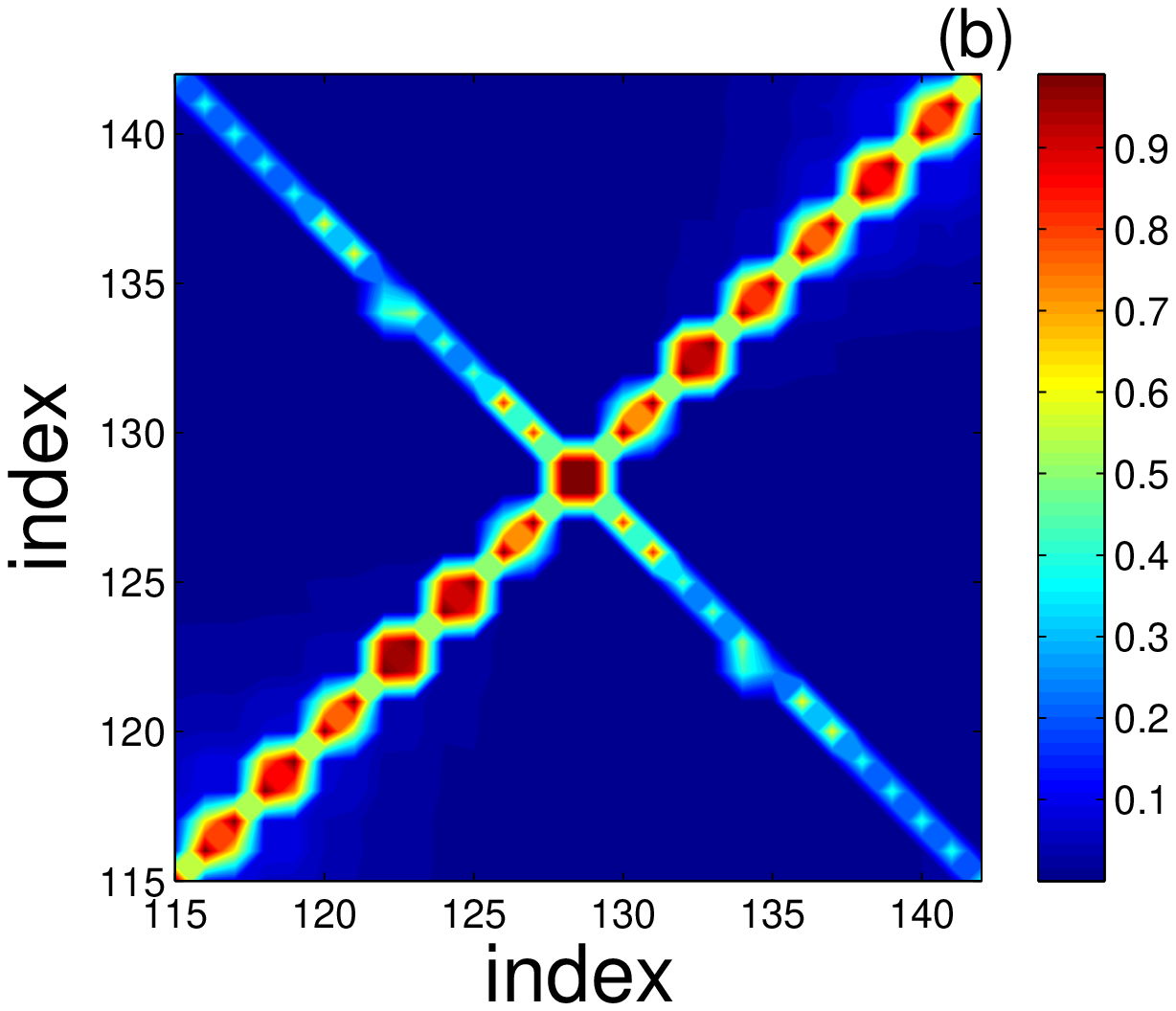}
\caption{\label{fig:f11}
(a) Contour plot of $\langle \cos(\theta) \rangle$ of CLVs for the Hamiltonian case. The local map is the skewed tent map with
$\epsilon=1.3$ and $r=0.15$. Panel (b) shows the enlargement of the central part of (a). Note that in the regime $\lambda \sim 0$ CLVs are nearly mutual orthogonal besides that
the conjugate pair of CLVs have a very small angle.
}
\end{figure}

Besides the mentioned symmetry resulting from the general Hamiltonian property, the conjugate pair of CLVs in systems with continuous symmetries such as 
Eq. (\ref{map-standard}) have some unexpected interesting features. The angle $\theta$ between a pair of CLVs is used to characterize their
relation, with $\cos(\theta)\equiv |\bf{e}^{(\alpha)} \cdot \bf{e}^{(\beta)}| $. The contour plot of the quantity $\langle \cos(\theta) \rangle$ 
is shown in Fig. \ref{fig:f11}, were $\langle \cdots \rangle$ means an average over time.

\begin{figure}
\includegraphics*[scale=0.3]{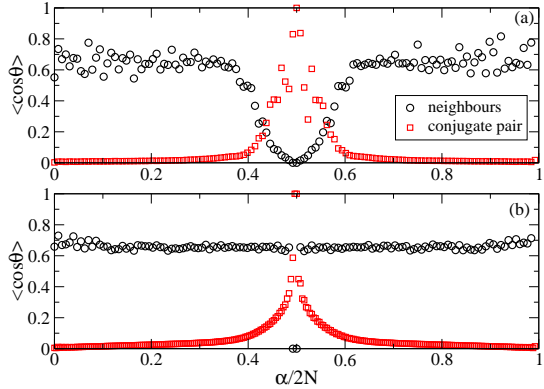}
\caption{\label{fig:f12}
The quantity $\langle \cos(\theta) \rangle$ of angles between the conjugate pair (square) and neighbouring CLVs (circle) of the Hamiltonian
system. The local map is (a) the skewed tent map with $r=0.15$ and (b) the sinusoidal map. The coupling strength is $\epsilon=1.3$. Switching to the nonhyperbolic case
 the orthogonality
between neighbours CLVs in the regime $\lambda \sim 0$ is broken.
}
\end{figure}

It shows that for the highly hyperbolic cases, for instance Eq. (\ref{map-standard}) with the special skewed tent map as the local dynamics, 
CLVs corresponding to near-zero Lyapunov
exponents are nearly orthogonal to each other as expected. The fact is more evident in Fig. \ref{fig:f12}. 
The near orthogonal nature of those CLVs explains the
observed similarity between CLVs and OLVs in Fig. \ref{fig:f7} and \ref{fig:f2} for the current parameter setting. 
In contrast the conjugate pair of CLVs tend to the same orientation as
approaching the zero Lyapunov exponents. With changing the local dynamics to the sinusoidal map 
the near orthogonal regime disappears completely whereas the qualitative behavior of the angle between conjugate pairs is hardly influenced.  

\section{conclusion and discussion}

We have explored, by using simple models of coupled map lattices, the similarity and difference between CLVs and OLVs, especially with respect to hydrodynamic Lyapunov
modes. For both Hamiltonian and dissipative cases, two different local maps were used to represent the typical situations with different degree of
hyperbolicity.
 The dynamics of the case with the special skewed tent
map is highly hyperbolic while the one with sinusoidal map is nonhyperbolic as most systems. 
In some sense the Hamiltonian system with the two local maps are corresponding to the
often used hard-core system and soft potential system respectively. 
With the replacement of OLVs by CLVs the formerly detected long wave-length structure in Lyapunov vectors can be seen as well.
Moreover the CLV $\lambda$-$k$ dispersion relation is linear for Hamiltonian system while quadratic for dissipative system as found for OLVs. The significance of HLMs as measured by
the static LV structure factor changes differently for Hamiltonian and dissipative systems with the replacement of OLVs by CLVs. For Hamiltonian systems the significance of HLMs
is always comparable for CLVs and OLVs independent of the variation of hyperbolicity as changing the local maps, besides that 
the OLVs with the most significant wave structure lie slightly away from the spectral center. Increasing system size tends to shift them back to the center.
 For dissipative systems the significance of HLMs is almost the
same for CLVs and OLVs if the special skewed tent map is used as local map. Departing from such a highly hyperbolic situation the HLM significance of CLVs reduces much faster
than that of OLVs.   

In the past there were already discussions regarding the symmetry of the conjugate pair of Lyapunov vectors in Hamiltonian system. It was found that owing to the symplectic
feature of Hamiltonian systems the coordinate and momentum
parts exchange their position for a conjugate pair of OLVs. A different symmetry is observed, however, for CLVs, namely that two CLVs in one conjugate pair are statistically
indistinguishable. As discussed the physical origin of this seemingly unreasonable property is the microscopic reversibility, a general feature of Hamiltonian systems.
For the specific issue of HLMs, it implies that the variation of HLM significance for CLVs is symmetric with respect to the spectral center. Besides that we found for CLVs that the
HLM significance is much lower in the momentum part than in the coordinate part. 

For the highly hyperbolic cases with the special skewed tent map as the local dynamics CLVs behave very similar to OLVs as demonstrated by the position and height of the
dominant peak of static LV structure factors. A direct monitoring of the mutual angle between CLVs shows that for those corresponding to near-zero Lyapunov exponents
the mutual angles are large and close to $\pi/2$. An unexpected observation is that the angle between conjugate pair decreases to zero as approaching the spectral center. Such a
feature persists as weakening the hyperbolicity. 

It is known that a dynamical system has two sets of OLVs, backward and forward ones. 
Only the backward OLVs, which can be calculated numerically via the standard method, are discussed in the main text. 
Similar results are expected for the forward OLVs except that they bear the
similarity to a different part of CLVs compared to the backward OLVs. A related discussion can be seen in the appendix \ref{ap3}.

To mention that a comparison of CLVs and OLVs in systems with hard-core interactions is performed by Posch et al \cite{posch-new}, 
which and the current contribution form a complementary view of the topic 
to each other.

\begin{acknowledgments}
This work is partially motivated by a question posed by an anonymous referee of our computing-time application to J\"ulich Supercomputing Centre.
Two appendices are motivated by a challenging discussion with Antonio Politi and Arkady Pikovsky during a workshop on Lyapunov analysis held in Florence at 2007.
We acknowledge discussions with Hugues Chat\'e, Arkady Pikovshy, Antonio Politi and Harald Posch and the financial support from the Deutsche Forschungsgemeinschaft (DFG Grant No. Ra416/6-1). 
\end{acknowledgments}

\appendix
\section{asymptotic and finite time Lyapunov exponents}
As can be seen from the definition and calculation algorithm in Sec. \ref{sec:algorithm} as well as from other sections CLVs and OLVs are
different in many respects. In this appendix we would like to point out that the (asymptotic) Lyapunov exponents corresponding to CLVs and OLVs
are identical but finite-time Lyapunov exponents (FTLEs) corresponding to these two sets of Lyapunov vectors are different in general.

From the calculation algorithm we know that a $k$-dimensional vector spanned $k$ arbitrary offset vectors will approach asymptotically the most
unstable $k$-dimensional subspace, which can be spanned by $k$ Lyapunov vectors, either CLVs or OLVs, 
associated with the first $k$ largest Lyapunov exponents. 
The growth rate $\Sigma_k$ of the volume $V_k$ of this $k$-dimensional subspace can be written as
\begin{equation}
\label{eq:le-clv}
\Sigma_k(t_1,t_2)=\sum_{i=0}^{k-1} \lambda_{(C)}^i(t_1,t_2)+\frac{1}{t_2-t_1}\ln | \frac{\prod_{i=1}^{k-1}\cos\theta^i(t_2)}{\prod_{i=1}^{k-1}\cos\theta^i(t_1)}  | 
\end{equation}
where $\lambda_{(C)}^i(t_1,t_2)$ is the growth rate of offset vectors along the $i$-th CLV, i.e. the $i$-th FTLE corresponding to this CLV and 
$\theta^i(t)$ is the angle between the $i$-th CLV and the subspace spanned by CLVs with index from $0$ to $i-1$.
Taking into account the mutual orthogonal nature of OLVs, the growth rate $\Sigma_k$ can be expressed as well by using characteristics of OLVs as
\begin{equation}
\label{eq:le-olv}
\Sigma_k(t_1,t_2)=\sum_{i=0}^{k-1} \lambda_{(O)}^i(t_1,t_2)
\end{equation}
where $\lambda_{(O)}^i(t_1,t_2)$ is the growth rate of offset vectors along the $i$-th OLV, i.e. the $i$-th FTLE corresponding to this OLV.

Combining Eq.(\ref{eq:le-clv}) and (\ref{eq:le-olv}) yields a simple relation between FTLEs
\begin{subequations}
\begin{equation}
\label{eq:ftle0}
\lambda_{(C)}^0(t_1,t_2)=\lambda_{(O)}^0(t_1,t_2)
\end{equation}
\begin{equation}
\label{eq:ftle}
\lambda_{(C)}^i(t_1,t_2)=\lambda_{(O)}^i(t_1,+t_2)+\frac{1}{t_2-t_1}\ln |\frac{\cos\theta^i(t_2)}{\cos\theta^i(t_1)}  |\,\,\,\,\text{for $i \in [1,N-1]$}
\end{equation}
\end{subequations}
with $N$ the dimension of the considered system. Since CLVs are in general not mutually orthogonal it is obvious from Eq. (\ref{eq:ftle}) that FTLEs 
$\lambda_{(O)}^i(t_1,t_2)$ and $\lambda_{(C)}^i(t_1,t_2)$ with $i\geq 1$ are normally different. 

As approaching the limit $t_2-t_1=+\infty$ the contribution of the second term in r.h.s. of Eq. (\ref{eq:ftle}) becomes negligible 
since the value
of $\cos \theta$ is bounded. This implies
\begin{equation}
\lambda_{(C)}^i(t_1,+\infty)=\lambda_{(O)}^i(t_1,+\infty)\,\,\,\,\text{for any $i \in [0,N-1]$},
\end{equation}
i.e. asymptotic Lyapunov exponents corresponding to CLVs and OLVs are identical.

\section{transformation properties of Lyapunov exponents, CLVs and hyperbolicity}
\subsection{Invariance of Lyapunov exponents and covariance of CLVs}
We consider a dynamical system which is written as 
\begin{equation}
\dot {\bf x}= {\bf F(x)} \text{        or         }  {\bf x}_{t+1}= {\bf F(x}_t{\bf)}.
\end{equation}
The time evolution of its trajectory can be expressed as
\begin{equation}
{\bf x}(t_2)= {\bf \phi}(t_1,t_2){\bf x}(t_1).
\end{equation}
Correspondingly the evolution of an infinitesimal perturbation vector with respect to the reference trajectory ${\bf x}(t)$ can be written as
\begin{equation}
\label{eq:linx}
\delta {\bf x}(t_2)= {\bf M}(t_1,t_2)\delta {\bf x}(t_1)
\end{equation}
with ${\bf M}=\partial \phi / \partial {\bf x}$.
Under a variable transformation ${\bf T: x \mapsto y}$ with
\begin{equation}
\label{tran-t}
{\bf y}= {\bf T}({\bf x}),
\end{equation}
the governing equation of infinitesimal perturbations becomes
\begin{equation}
\label{eq:liny}
\delta {\bf y}(t_2)= {\bf M'}(t_1,t_2)\delta {\bf y}(t_1).
\end{equation}
Here the two variables $\delta {\bf y}$ and $\delta {\bf x}$ are related via a linear transformation ${\bf L: \delta x \mapsto \delta y}$ with
\begin{equation}
\label{tran-l}
\delta{\bf y}= {\bf L}\delta{\bf x}
\end{equation}
It is known that the linear transformation ${\bf L }$ is determined by the transformation ${\bf T}$ via
\begin{equation}
\label{eq:lt}
{\bf L}= {\bf D_x T}({\bf x}),
\end{equation}
where $({\bf D_x T})_{ij}=\partial T_i/\partial x_j$.
By using Eq.(\ref{eq:lt}), (\ref{eq:linx}) and (\ref{eq:liny}) one can show that
\begin{equation}
\label{eq:mt1}
{\bf M'L}= {\bf LM}.
\end{equation}
which means that ${\bf M'}$ and ${\bf M}$ are related via a similarity transformation ${\bf M'}= {\bf LML^{-1}}$ if ${\bf L}$ is invertible.

If ${\bf e(t_1)\equiv e(x}(t_1)) $ is a CLV in the ${\bf x}$-coordinate system, it satisfies the condition
\begin{equation}
\label{eq:clvx}
{\bf M}(t_1,t_2) {\bf e}(t_1)=\sigma(t_1,t_2) {\bf e}(t_2)
\end{equation}
with $\lambda=\lim_{t_2-t_1 \to\infty }\frac{1}{t_2-t_1}\ln|\sigma(t_1,t_2)|$ being the Lyapunov exponent corresponding to this CLV.
Multiplying the both sides of Eq.(\ref{eq:clvx}) with ${\bf L}$ and using Eq.(\ref{eq:mt1}) results in
\begin{equation}
\label{eq:clvy}
{\bf M'}(t_1,t_2) {\bf L e}(t_1)=\sigma(t_1,t_2) {\bf L e}(t_2).
\end{equation}
Denoting ${\bf e'}(t)={\bf L e}(t)/\|{\bf L e}(t)\|$ one can reformulate Eq.(\ref{eq:clvy}) as
\begin{equation}
{\bf M'}(t_1,t_2) {\bf e'}(t_1)=\sigma(t_1,t_2) \frac{\|{\bf L e}(t_2)\|}{\|{\bf L e}(t_1)\|}{\bf  e'}(t_2)=\sigma'(t_1,t_2){\bf  e'}(t_2).
\end{equation}
Under the condition that
\begin{equation}
\label{eq:clvy3}
\lim_{t_2-t_1 \to\infty }\frac{1}{t_2-t_1} \ln(\frac{\|{\bf L e}(t_2)\|}{\|{\bf L e}(t_1)\|})=0
\end{equation}
one can easily obtain that 
\begin{equation}
\label{eq:clvy2}
\lim_{t_2-t_1 \to\infty }\frac{1}{t_2-t_1} \ln | \sigma'(t_1,t_2) |=\lim_{t_2-t_1 \to\infty} \frac{1}{t_2-t_1} \ln |\sigma(t_1,t_2)|,
\end{equation}
which implies that (i) the unit vector ${\bf e}'(t)$ is a CLV in the ${\bf y}$-coordinate system and it is related to ${\bf e}(t)$ via ${\bf e'}(t)={\bf L e}(t)/\|{\bf L e}(t)\|$;
 (ii) the asymptotic Lyapunov exponent associated with ${\bf e}'(t)$
is identical to the asymptotic Lyapunov exponent corresponding to ${\bf e}(t)$; (iii) the finite-time Lyapunov exponent 
$\lambda'(t_1,t_2)=\frac{1}{t_2-t_1} (\ln|\sigma(t_1,t_2)|+\ln \frac{\|{\bf L e}(t_2)\|}{\|{\bf L e}(t_1)\|})$ in the ${\bf y}$-coordinate system is different
from the one $\lambda(t_1,t_2)=\frac{1}{t_2-t_1} \ln|\sigma(t_1,t_2)|$ in the ${\bf x}$-coordinate system.

\begin{figure}
\includegraphics*[scale=0.3]{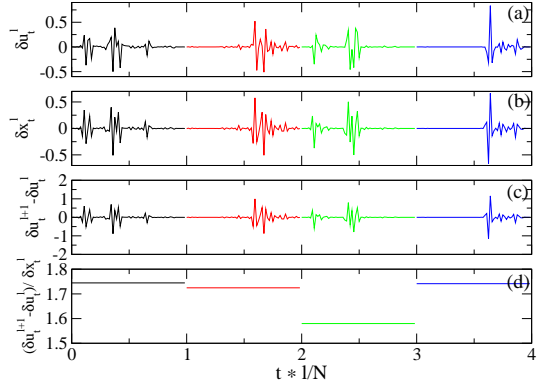}
\caption{\label{fig:ap-clv}
Instantaneous profiles of CLVs of two systems related via variable transformation Eq.(\ref{tran}). As shown in panel (d) CLVs of
the two systems are related via the relation given in Eq.(\ref{u-relation}).
}
\end{figure}

\begin{figure}
\includegraphics*[scale=0.3]{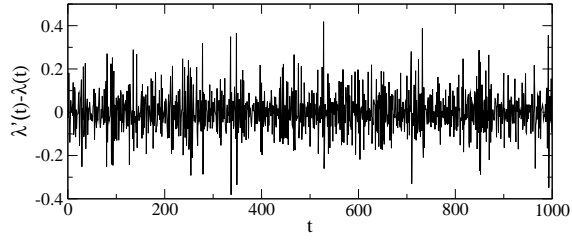}
\caption{\label{fig:ap-lep}
Difference between finite-time Lyapunov exponents of two systems related via variable transformation Eq.(\ref{tran}).
}
\end{figure}

For an invertible transformation ${\bf T}$ with the assumption that the reference trajectory is bounded in phase space one can easily show the boundedness of ${\bf L}$
\cite{eichhorn}, i.e.
\begin{equation}
L^-\| {\bf  e}\| \leq \|{\bf L e}  \| \leq L^+\| {\bf  e}\| 
\end{equation}
for two constant $L^-\leq L^+<\infty $, which implies the validness of the condition stated in Eq.(\ref{eq:clvy3}). As discussed in \cite{eichhorn} these requirements on ${\bf T}$ can be weakened such that for
non-invertible transformations ${\bf T}$ one can still get the invariance of Lyapunov exponents and the covariant transformation of CLVs. This is also confirmed by our numerical example
below.

As discussed already in Ref.\cite{hlm-cml}, via the transformation 
\begin{equation}
\label{tran}
x_t^l=u_t^{l+1}-u_t^l
\end{equation}
Eq.(\ref{map-circle}) can be mapped to the following diffusively coupled CMLs
\begin{equation}
\label{map-bohr}
x_{t+1}^l=x_t^l+\epsilon [ f(x_t^{l+1})+f(x_t^{l-1})-2f(x_t^l)].
\end{equation}
According to our above arguments CLVs of the two system are related via the transformation
\begin{equation}
\label{u-relation}
\delta x_{t}^{(\alpha)l}=c_t(\delta u_t^{(\alpha)l+1}-\delta u_t^{(\alpha)l})
\end{equation}
where $c_t$ is a time-dependent normalization factor. Numerical results shown in Fig. \ref{fig:ap-clv} and \ref{fig:ap-lep} for a case with the sinusoidal map $f(z)$ confirms our conclusions.
Note that the transformation Eq.(\ref{tran}) is non-invertible.

\subsection{Invariance of hyperbolicity under diffeomorphisms}

In viewing that ${\bf e'}(t)={\bf L e}(t)/\|{\bf L e}(t)\|$ one would expect that the absolute value of angles between CLVs is not invariant under the variable transformation. 
Whether the angle is zero or not, i.e the feature of hyperbolicity, is expected to be preserved under diffeomorphisms. This conjecture is supported by the following arguments.

If the variable transformation ${\bf T}$ given in Eq.(\ref{tran-t}) is a diffeomorphism, the corresponding transformation ${\bf L}$ of the perturbation in Eq.(\ref{tran-t}) would be an invertible linear transformation.

Consider two CLVs ${\bf  e}_1(t)$ and ${\bf  e}_2(t)$ in the $x$-coordinate system and denote the corresponding CLVs in 
the $y$-coordinate system as ${\bf  e}'_1(t)$ and ${\bf  e}'_2(t)$. Since an affine transformation like $ {\bf L }$ preserves the collinearity of points the angle
 between the transformed CLVs ${\bf  e}'_1(t)$ and ${\bf  e}'_2(t)$ is zero if the angle between original CLVs ${\bf  e}_1(t)$ and ${\bf  e}_2(t)$ is zero, 
 i.e. $\angle({\bf  e}_1(t),{\bf  e}_2(t))=0$ implies $\angle({\bf  e}'_1(t),{\bf  e}'_2(t))=0$. 
 Similar arguments for the inverse ${\bf L}^{-1}$ leads to that $\angle({\bf  e}'_1(t),{\bf  e}'_2(t))=0$ implies $\angle({\bf  e}_1(t),{\bf  e}_2(t))=0$. These properties indicate the
 preservation of the collinearity of CLVs under diffeomorphisms. 
 
Consider now two subspaces $S_1$ and $S_2$ spanned by two sets of different CLVs $\{ {\bf e}_{S_1}^i\}$ and $\{ {\bf e}_{S_2}^i\}$. 
If the angle between the two subspaces is zero in one coordinate system it means that the two sets of CLVs are linearly dependent, i.e.
$\sum_{i}c_i{\bf e}_{S_1}^i +d_i{\bf e}_{S_2}^i=0$ for certain constants $c_i$ and $d_i$. Preservation of collinarity under affine transformations implies
 that the corresponding transformed CLVs
are linearly dependent as well, i.e. the angle between subspaces in the transformed coordinate system is also zero.
 Similarly one can show that if the angle between two subspaces is nonzero in one coordinate system it would be nonzero in other transformed
coordinate systems, too. These arguments show that whether the angle between subspaces is zero or not is invariant under dffeomorphisms, i.e. the property of hyperbolicity is preserved. 
 
Similar to the discussion about the transformation properties of CLVs one can weaken the requirements on the transform ${\bf T}$ but rather the same conclusion about the
hyperbolicity can be reached.
A known example is the Kuromato-Sivashinsky equation. It can be written in two different forms as
\begin{equation}
u_t=u_{xx}+u_{xxxx}+u_x^2/2
\end{equation}
or 
\begin{equation}
v_t=v_{xx}+v_{xxxx}+vv_x
\end{equation}
which are related via a non-invertible transformation $v=u_x$.
Numerical simulations show that the two forms have the identical hyperbolicity and details will be
shown elsewhere \cite{yang2}. 

\section{analytical calculation of CLVs and OLVs of a Hamiltonian system}
\label{ap3}
For the Hamiltonian system Eq.(\ref{map-standard}) with the limiting case $r=0$ of the skewed tent map Eq.(\ref{map-sktent}) one can calculate the CLVs and OLVs analytically. Consistence with 
numerical results presented in the main part of the paper can thus be checked.  

\subsection{CLVs}

For the case $r=0$ in Eqs.(\ref{map-standard},\ref{map-sktent}) the time evolution of the infinitesimal perturbations is governed by 
 \begin{equation}\delta \vec{\Gamma}_{t+1}=\left( 
 \begin{matrix}
 I_L+\epsilon D_L & I_L\\ \epsilon D_L & I_L
 \end{matrix} \right)
 \cdot \delta \vec{\Gamma}_{t} \end{equation}
  where $\delta \vec{\Gamma}_{t} \equiv \{ \delta u_t^1, \delta
 u_t^2, \cdots, \delta u_t^L;\delta v_t^1, \delta v_t^2, \cdots, \delta v_t^L \}$ is the offset
 vector in the tangent space and $I_L$, $D_L$ denote the $(L\times L)$-unit matrix and 
the discrete Laplacian, respectively.
 Notice that the fundamental matrix 
\begin{equation}
M_2 \equiv \left( 
 \begin{matrix}
 I_L+\epsilon D_L & I_L\\ \epsilon D_L & I_L
 \end{matrix} 
  \right)
 \end{equation} 
 is time independent and thus the eigenvectors of $M_2$ are CLVs of this system.  

By using
 the eigenvectors $\vec{e}^{(\alpha)}$ of the matrix $D_L$
the eigenvectors of the fundamental matrix 
 $M_2$ can be constructed as $\{ \vec{e}^{(\alpha)};c(k)\vec{e}^{(\alpha)} \}$. The associated eigenvalues are
  \begin{equation}
  \label{eigenvalue}
\mu_{\pm}(k)= \frac{\eta(k)+2\pm \sqrt{\eta^2(k)+4\eta(k)}}{2}
\end{equation} 
where $\eta(k)=-2\epsilon (1-\cos k)$ and
 the corresponding $c(k)$ can be calculated as
 \begin{equation}
 \label{ck}
c_{\pm}(k)= \mu_{\pm}(k)-1-\eta(k).
\end{equation}
The following properties of these eigenvectors/CLVs can be obtained:

i) The corresponding eigenvalues satisfy $\mu_{+}(k)\mu_{-}(k)=1$ which indicates the conjugate pair property $\lambda_+=-\lambda_-$ of Lyapunov exponents since $\lambda \equiv \ln| \mu(k)|$.

ii) The group of CLVs $\{ \vec{e}^{(\alpha)};c_+(k)\vec{e}^{(\alpha)} \}$ corresponding to positive Lyapunov exponents are in general not orthogonal to CLVs $\{ \vec{e}^{(\alpha)};c_-(k)\vec{e}^{(\alpha)} \}$ corresponding to the negative branch of the Lyapunov spectrum,
although members of either group are mutually orthogonal. This indicates that these
eigenvectors/CLVs are not OLVs of that system.

iii) As approaching the spectral center $\lambda=0$, one has $k\to 0$. Two CLVs in a conjugate pair tend to be collinear, i.e. $\cos \theta \to 1$, where $\theta$ denotes the angle between that pair of
CLVs. More precisely, as $k\to 0$, one has 
  \begin{equation}
\eta(k) \approx \epsilon k^2, 
\end{equation} 
  \begin{equation}
  \mu_{\pm}(k) \approx 1\pm  \sqrt{\epsilon}k
\end{equation} 
and
  \begin{equation}
  c_{\pm}(k) \approx \pm  \sqrt{\epsilon}k.
\end{equation} 
Thus for a conjugate pair of CLVs $\{ \vec{e}^{(\alpha)};\sqrt{\epsilon}k\vec{e}^{(\alpha)} \}$ and $\{ \vec{e}^{(\alpha)};-\sqrt{\epsilon}k\vec{e}^{(\alpha)} \}$ 
one has
  \begin{equation}
  \cos \theta \approx 1-\epsilon k^2.
\end{equation} 
In consistence to this, we reported in Sec. \ref{sec:pair} that for the cases with $r$ close to 0 as shown in Fig. \ref{fig:f11} and \ref{fig:f12} neighbouring CLVs are nearly orthogonal
while those in a conjugate pair tend to be collinear as approaching the spectral center $\lambda=0$.

\subsection{OLVs}

Now we start to calculate OLVs of this system, which are eigenvectors of the matrix $M_2^n(M_2^T)^n$ as $n$ goes to infinity, where $M_2^T$ is the transpose of $M_2$.

The matrix $M_2$ has the similar transformation $M_2=Q\Lambda Q^{-1}$, where the column vectors of $Q$ are eigenvectors of $M_2$ and entries of the diagonal matrix $\Lambda$ are corresponding
eigenvalues $\mu_{\pm}(k)$ mentioned above.
The matrix $M_2^n(M_2^T)^n$ can thus be written as
  \begin{equation}
M_2^n(M_2^T)^n=Q \Lambda^n Q^{-1} (Q^{-1})^T \Lambda^n Q^T. 
\end{equation} 
Considering the orthogonal nature of $\vec{e}^{(\alpha)}$, the discussion of eigenvalue and eigenvectors can be simplified by using
 the submatrix $Q(k)$ and $S(k) \equiv M_2^n(M_2^T)^n(k)$ related to the
vectors $\vec{e}^{(\alpha)}(k)$.
They are
  \begin{equation}
Q(k)=
\left( 
 \begin{matrix}
 1 & 1\\ c_+(k) & c_-(k)
 \end{matrix} 
  \right),
\end{equation} and
  \begin{equation}
S(k)=
(c_+(k)-c_-(k))^{-2}\left( \begin{matrix}
 a(k) & d(k)\\ d(k) & b(k)
 \end{matrix} \right),
\end{equation}  
where $a=\mu_+^{2n}(1+c_-^2)+2\mu_+^n\mu_-^n+\mu_-^{2n}(1+c_+^2)$, $b=c_+^2\mu_+^{2n}(1+c_-^2)+2c_+c_-\mu_+^n\mu_-^n+c_-^2\mu_-^{2n}(1+c_+^2)$, 
and $d=c_+\mu_+^{2n}(1+c_-^2)+(c_++c_-)\mu_+^n\mu_-^n+c_-\mu_-^{2n}(1+c_+^2)$.
The eigenvalues of the matrix $S(k)$ can be obtained as 
  \begin{equation}
  \xi_{\pm}(k) = \frac{a+b\pm \sqrt{(a+b)^2-4(ab-c^2)}}{2}.
\end{equation} 
As the eigenvectors of the matrix $M_2^n(M_2^T)^n$ can be constructed as $\{ \vec{e}^{(\alpha)};p\vec{e}^{(\alpha)} \}$, one can easily get
  \begin{equation}
  p(k)=\frac{\lambda-a}{d}=\frac{(b-a)/2+\sqrt{(a+b)^2-4(ab-c^2)}}{d}.
\end{equation} 
Since $\mu_{+}(k)>1>\mu_{-}(k)$, as $n$ goes to infinity one has $p(k) \to c_+(k)$ with the corresponding $\lambda_+(k) =\frac{1}{n}\ln |\xi_+(k)| \to \ln| \mu(k)|$, which indicates that the OLVs associated with positive
Lyapunov exponents are the same as the corresponding CLVs.

In consistence to this, as shown in Fig. \ref{fig:f2}d and \ref{fig:f4a}a, for cases with $r$ close to $0$ OLVs and CLVs associated with positive Lyapunov
exponents are very similar.

A dynamical system has actually two sets of OLVs, namely backward and forward OLVs, which are eigenvectors of the matrix $M_2^n(M_2^T)^n$ and $(M_2^T)^nM_2^n$ as $n$ goes to infinity,
respectively. In above discussions the backward OLVs are used since they are the ones numerically obtained from the standard method \cite{benettin}.
 For the forward OLVs one can do the
similar calculations and the conclusion is that a half of the forward OLVs are the same as the CLVs associated with negative Lyapunov exponents.


\begin{thebibliography}{50}
\bibitem{note-ergocicity}For those Hamiltonian systems which are nonergodic one can always find a decomposition of the invariant measure 
to ergodic components and we
expect that the two trajectories talked are in the same ergodic component.
\end{thebibliography}
\end{document}